# Migration of helium-pair in metals

J. L. Cao and W. T. Geng[1]

*School of Materials Science & Engineering, University of Science & Technology Beijing, Beijing 100083, China*

## Abstract

Understanding helium accumulation in plasma-facing or structural materials in a fusion reactor starts from uncovering the details of the migration of single and paired He interstitials. We have carried out a first-principles density functional theory investigation into the migration of both a single interstitial He atom and an interstitial He-pair in *bcc* (Fe, Mo and W) and *fcc* (Cu, Pd and Pt) metals. By identifying the most stable configurations of an interstitial He-pair in each metal and decomposing its motion into rotational, translational, and rotational-translational routines, we are able to determine its migration barrier and trajectory. Our first-principles calculations reveal that the migration trajectories and barriers are determined predominantly by the relatively stable He-pair configurations which depend mainly on the stability of a single He in different interstices. Contrary to atomistic studies reported in literature, the migration barrier in *bcc* Fe, Mo, and W is 0.07, 0.07, and 0.08 eV respectively, always slightly higher than for a single interstitial He (0.06 eV for all three). Configurations of a He-pair in *fcc* metals are much more complicated, due to the stability closeness of different interstitial sites for a single He atom. In both Cu and Pd, the migration of a He-pair proceeds by moving one He at a time from one tetrahedral site to neighboring octahedral site; whereas in Pt the two He move simultaneously because the bridge interstitial site presents an extremely low barrier. The migration barrier for a He-pair is 0.05, 0.15, and 0.04 eV for Cu, Pd, and Pt, slightly lower than (in Cu), or similar to (in Pd and Pt) a single He, which is 0.08, 0.15, and 0.03 eV, respectively. The associative motions of a He-pair are ensured by the strong He-He interactions in metals which are *chemical bonding*-like and can be described very well with Morse potentials.

---

[1]E-mail: geng@ustb.edu.cn

## I. INTRODUCTION

In a working fusion reactor, large amount of helium (He) are produced from transmutation reactions in the structural materials or introduced from the edge plasma into the plasma facing materials. Because of the close-shell electronic structure, He atoms in solids have a strong tendency to segregate into the low charge-density region and to bind strongly with vacancies,[1-3] grain boundaries,[4-8] dislocation cores,[9,10] precipitates[11,12] and impurity atoms[13-16] and would induce bubble formation and void swelling, which in turn result in high temperature embrittlement, surface roughening, and blistering.[17-20]

Although the evolution of He in metals has been studied for over 50 years, our understanding of the He behavior in metals is still far from complete.[21] Experimental techniques such as thermal desorption and TEM after He irradiation or implantation can yield some information about He positioning. However, the formation, migration, and growth of He bubbles and their interaction with other defects, are difficult to observe directly in experiment.[22,23] On the other hand, theorists also encounter fundamental challenges in attempting to describe the evolution of a He bubble under interaction with the surrounding crystal lattice. The positioning of a single He atom and its migration have now been well studied, thanks to the predictive power of density functional theory (DFT) calculations.[3,24,25] First-principles computational method is now capable of calculating the solubility and diffusivity of foreign interstitial atoms in metals at a level of accuracy close to and sometimes better than available from experiments.[26] A fully quantum-mechanical treatment of the migration of a He cluster in solid, nevertheless, is extremely demanding due to the large number of structural freedoms of the cluster. To make the computation affordable, less precise approaches such as Molecular dynamics (MD) simulations employing empirical potentials and Monte Carlo method have often been used.[27-29]. Good knowledge of the behavior of small He clusters serving as bubble embryos is highly imperative without which the details of the He bubble growth would be out of reach.

The migration of small He clusters in Fe, a candidate for structural material in fusion reactor, and W, a candidate for plasma-facing material, have been intensively studied by MD simulations using various meticulously constructed inter-atomic potentials and parameters.[27, 28, 30-32] But unfortunately, the calculated values of migration barrier from different inter-atomic potentials show remarkable discrepancies, casting some doubt on the accuracy of predictions given by these simulations. This is a strong indication, in our view of point, that the construction of efficient inter-atomic potentials should involve the migration barrier not only for a single He atom, but also that for a He pair or even a He trimmer which can now only be determined by first-principles DFT calculations, but not experiment.

In a recent work, we have attempted to evaluate the migration barrier for a He-pair in W using first-principles DFT calculations.[33] Our approach to determining the trajectory of an interstitial He pair moving in W is composed of two steps. First, to identify the most stable configurations of a He pair in W, which are presumably the local energy minima on the migration path. And second, to decompose the motion of a He pair into rotational, translational, and rotational-translational routines. By connecting these routines on the probable path, we were able to determine migration barrier and trajectory of a He pair. Now, we attempt to extend our study on He-pair to more metallic systems and try to uncover the rules underlying its migration behavior. The strong attraction of interstitial He atoms in most metals make it easy to form interstitial He-pair or clusters even in absence of vacancies or other crystalline defects.[34] Therefore, the specific numerical results should be of great interest for researchers developing the inter-atomic potentials. And more importantly, through a systematic investigation, we may expect to discover some general rules governing the migration of He-pairs, which will definitely be relevant to understand the migration of larger clusters.

The metals we choose include not only those with *bcc* lattices, but also some with *fcc*

structures. For *bcc* metals, we have studied Fe, a candidate for structural material in fusion reactor, and Mo, which is expected to show similar features to W concerning He migration. Since we have discovered a new low-barrier migration path in Fe and Mo, we also revisited the W system. As for *fcc* metals, we choose Cu, Pd, and Pt as model systems. Our interest in Cu and Pt is stimulated by an intriguing phenomenon, the formation of nanostructure (*fuzz*) on metal surface by He plasma irradiation.[35] Under similar irradiation conditions, *fuzz* was found on W, but not on Cu, Ti, stainless steel, and Pt. Platinum surface even retained metallic luster, while the surfaces of Ti and stainless steel turned black and the Cu surface layer was peeled on some parts.[35] Baldwin et al.[36] investigated the growth kinetics of *fuzz* in W in experiment, and found that the thickness of *fuzz* layer increases with time by $t^{1/2}$ dependence. Several models of the *fuzz* mechanism have been proposed, including viscoelasticity based on high−pressure He bubble,[37] stress-driven bubble growth,[38] and the knock-out W adatom formation[39]. A more sophisticated model developed by Nordlund et al.[40] argued that, instead of the mechanism mentioned above, it is the balance between loop punching and He bubble rupture that causes the kinetic surface roughening. With a number of assumptions such as spherical bubble growth, this model could reproduce the qualitative fuzz growth behavior and certain quantitative aspects demonstrated by experiments, such as the *fuzz* thickness growing as $t^{1/2}$. More recently, Sandoval *et al*. examined the growth of He bubble in W with the rate spanning 6 orders of magnitude under more realistic conditions, using MD and Parallel replica.[41] They revealed the crucial role played by the mobility of interstitials near the bubble surface. In both works,[40, 41] the kinetics of bubble growth and interstitial diffusion significantly influence the and morphology of the W surface. It appears that these models can give reasonable explanation for *fuzz* growth on W surface; however, whether they are applicable to other metals in general, i.e., to reflect distinct material-dependent properties, remains unclear. Apparently, to make high-quality large scale simulation on the surface roughing, precisely determined inputs parameters, migration barriers of small He clusters are urgently needed.

The remainder of this paper is organized as follows. In Sec. II, Methodology and computational details are described. Sec. III presents the calculations of migration barriers and paths of both single He and He-He pair. The underlying physics for migration is discussed in Sec. IV. Finally, a summary of conclusions is given in Sec. V.

## II. METHODOLOGY

The first-principles DFT calculations were performed using the Vienna *Ab initio* Simulation Package (VASP).[42] The electron-iron interaction was described using projector augmented wave (PAW) method[43] and the exchange and correlation were treated with generalized gradient approximation (GGA) in the Perdew Burke Ernzerhof (PBE) form.[44] The cut-off energy for the plan wave basic set was set to 480 eV which has been shown to be enough for the closed shell of He 1*s* electrons. For *bcc* and *fcc* metals, we employed 4×4×4 128-atom and 3×3×3 108-atom supercell, respectively. A 3×3×3 *k*-mesh in Monkhorst-Pack Scheme replaces the integration over Brillouin zone. For each system, geometry optimization would continue until the total energy of this system was converged to less than $10^{-4}$ eV·Å$^{-1}$. Nudged elastic band (NEB) method was adopted to calculate the migration barrier.[45] Five images were linearly interpolated between starting and terminal point on the migration track. In structure optimization and NEB calculations, the volume and shape of the supercells were fixed at the values for the corresponding clean system and the atomic positions were fully relaxed.

## III. NUMERIACL RESULTS

### A. Positioning and migration of a single interstitial He

An interstitial space (interstice) is a space between atoms in solid. The octahedral interstitial sites (OIS) and tetrahedral interstitial sites (TIS) are the two types of interstitial sites in *bcc* and *fcc* metals (Fig.1) where the electron density reaches a local minimum. It is now well recognized that a single He in *bcc* metals such as Fe, Mo, W, V, Nb, Ta will occupy a TIS.[24] The electron density at TIS is lower than OIS in *bcc* crystals, although its volume is smaller the latter. A TIS in bcc lattice has four nearest neighbors at a distance of 0.559*a* (*a* is the lattice constant) while an OIS has two nearest neighbors at a distance of 0.5*a* and four next-nearest-neighbors at 0.707*a*. It is the lesser separated two nearest-neighbors that make the He more compressed than in a TIS. This argument, however, cannot explain the stability of He in *fcc* metals. In *fcc* lattices, both TIS and OIS are in the center of a regular polyhedron. A TIS has four nearest neighbors at 0.433*a*, while an OIS has six near neighbors at 0.5*a*. One is prompted to expect that He will prefer OIS over TIS. Nevertheless, first-principles DFT calculations demonstrated that this is true for 4d and 5d transition metals like Pd, Ag, Pt, Au, but not 3d metals such as Ni and Cu.[25]

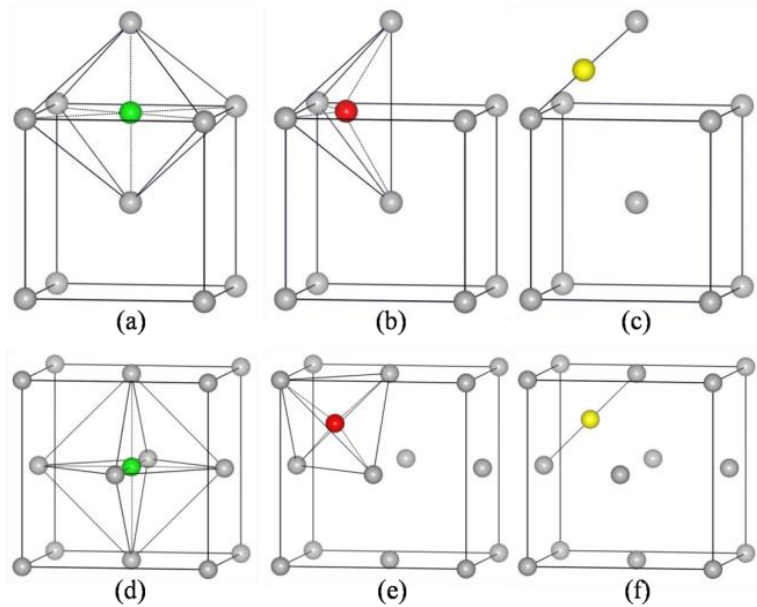

**Fig.1. The octahedral (a, d), tetrahedral (b, e), and bridge (c, f) interstitial sites (color spheres) in *bcc* (top) and *fcc* (bottom) lattices.**

A will be discussed in Sec. IV, we have figured out an approach to separate the two contributions to the embedding energy of He in TIS and OIS: one is the compression from a homogeneous electron gas, and the other from the compression imposed by surrounding matrix atoms, denoted by *point elastic constant*. In the case of Cu, although the electron density at TIS is higher than that at OIS, its effect can be over compensated by the lower *point elastic constant* at TIS than that at OIS. As a consequence, a He atom has a small embedding energy in TIS than in OIS. Inspired by the intriguing fact that an interstitial site with high electron density and small free volume could still possibly be preferred by He, we have extend our search of He positioning in metals to a less noticed interstitial site, i.e., the middle point between two nearest neighboring atoms. Here we denote it as bridge interstitial site (BIS). Obviously, in *fcc* lattices a BIS is just in between two nearest neighboring OIS or two nearest neighboring TIS.

**Table I. The calculated formation energy (eV) of an interstitial He in octahedral (OIS), tetrahedral (TIS), and bridge (BIS) sites in various *bcc* and *fcc* metals. Also listed is the calculated migration energy (eV) for an interstitial He in each metal. Results from literature are listed in parenthesis for comparison.**

| Matrix | *bcc* | | | *fcc* | | |
|---|---|---|---|---|---|---|
| | Fe | Mo | W | Cu | Pd | Pt |
| Lattice (Å) | 2.83 | 3.16 | 3.17 | 3.63 | 3.96 | 3.98 |
| $E_{OIS}^f$ | 4.81 | 5.53 | 6.34 | 4.02 | 3.60 | 4.83 |
| | (4.75[a]) | (5.48[a]) | (6.41[a]) | (3.82[a]) | (3.58[a], 3.68[b]) | (4.73[b]) |
| $E_{TIS}^f$ | 4.61 | 5.36 | 6.14 | 4.00 | 3.73 | 5.06 |
| | (4.56[a]) | (5.33[a]) | (6.19[a]) | (3.80[a]) | (3.70[a], 3.82[b]) | (5.18[b]) |
| $E_{BIS}^f$ | 5.01 | 5.83 | 6.69 | 4.13 | 3.72 | 4.86 |
| $E_{He1}^m$ | 0.06 | 0.06 | 0.06 | 0.08 | 0.15 | 0.035 |
| | (0.06[c]) | (0.06[d]) | (0.06[e]) | (0.07[f]) | | |

a-Ref.25, b-Ref.46, c-Ref.3, d-Ref.47, e-Ref.34, f-Ref.48.

The calculated formation energy of an interstitial He in OIS, TIS, and BIS in *bcc* Fe, Mo and W and *fcc* Cu, Pd and Pt is listed in Table I. The formation energy is defined as:

$$E_{int,M}^f = E(\text{He} + n\text{M}) - E(n\text{M}) - E(\text{He}) \quad (1)$$

Where $E(\text{He}+n\text{M})$ and $E(n\text{M})$ are the total energy of a supercell containing *n* M atoms with and without a He atom, and $E(\text{He})$ is the total energy of a free He, which is taken as zero. In agreement with reference 24, 25, TIS are always the most stable interstitial sites in *bcc* transition metals. The most stable interstitial site in Cu, Pd and Pt is TIS, OIS and OIS, respectively. Very interestingly, we find that in BIS, where the electron density does not reach a local minimum, but is rather a saddle point, He is nearly as stable as in a TIS (Pd) or even as in OIS (Pt). This strongly suggests that BIS has to be considered in our search of the stable configurations for a He-He pair in Pt.

As for the zero-point-energy (ZPE), we have performed this correction only for *fcc* systems, because in bcc metals, the stability of a He is remarkably different in different interstitial site and the ZPE was found not to be significant.[13] Our calculations show that although the ZPE correction for He is not negligible in Cu, Pd, and Pt, the relative stability of OIS, TIS, and BIS is essentially not changed.

To determine the migration path, we have always set the starting point as the most stable interstitial site. In the three *bcc* metals, the migration barrier for a single He is found to be the same within the numerical precision. It is as low as 0.06 eV, from a TIS to one of its neighboring TIS [Fig.2(a)]. These results are in excellent agreement with previous reports.[3, 34, 47] In Cu, the easiest path to migrate is not a direct TIS→TIS (0.15 eV), but rather a TIS→OIS→TIS (0.08 eV), as was discovered in reference.[48] The path in Pd with lowest energy barrier (0.15eV) is found to be a direct OIS→OIS. In Pt, due to the existence of low-energy BIS and large stability difference between OIS and TIS, a direct OIS→OIS migration experiences a barrier is found to be as low as 0.035 eV.

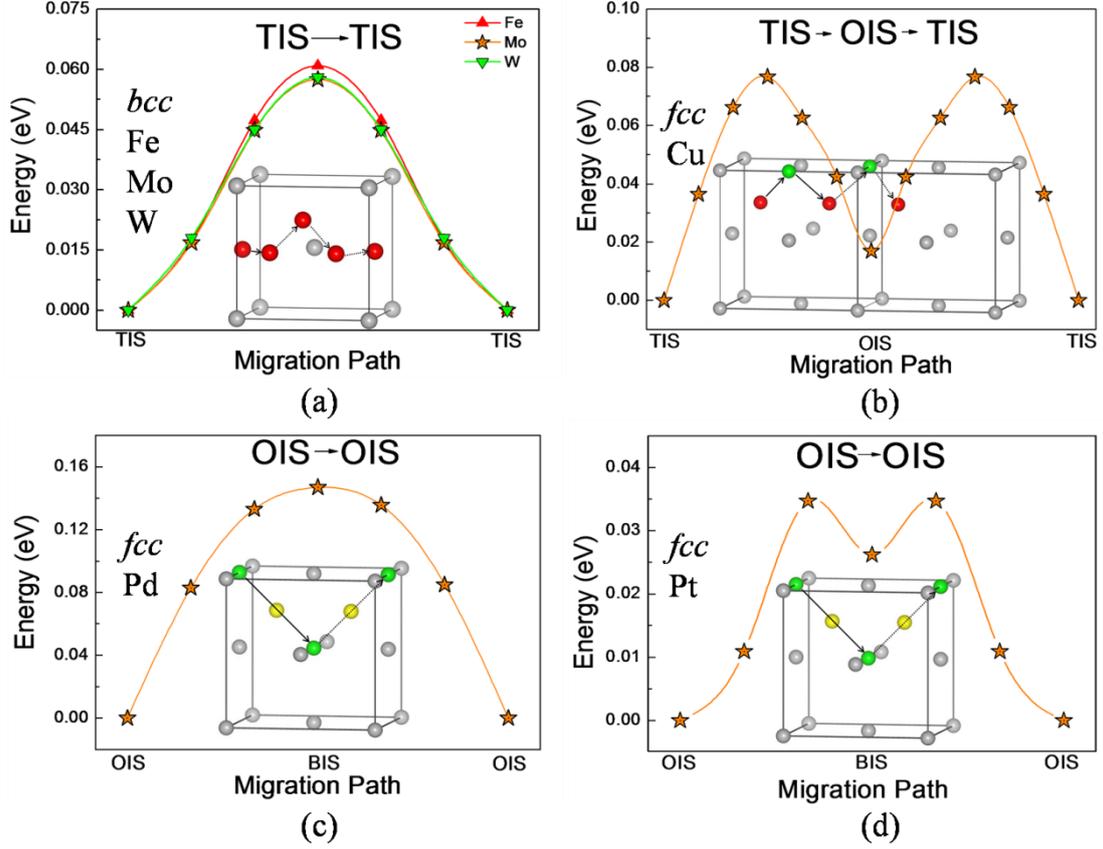

**Fig.2.** Migration energy of a single He in various metals: (a) Fe, Mo and W; (b) Cu; (c) Pd; (d) Pt. Migration trajectory is shown in inset. Green, red, yellow and gray spheres represent OIS, TIS, BIS sites and metal atoms.

### B. Stability of He-pair configurations

The binding energy of a pair of interstitial He atoms is defined as the energy cost when the two are separated to far apart, each in its most stable site:

$$E_{He2}^b = [2E(He + nM)] - [E(2He + nM) + E(mM)] \qquad (2)$$

$E(He+nM)$ and $E(2He+nM)$ are the total energy of the supercell containing $n$ metal atoms and one He or a He-He pair, respectively. We note that in the supercell containing one He, the He is in its most stable position. A Positive $E_{He2}^b$ means attraction. For a complete search for the most stable configurations of a He-pair, we have investigated three groups of combinations including TIS+TIS, OIS+OIS and OIS+TIS, each with varying separation distance, for all three *bcc* metals, Cu and Pd. Since BIS is almost as stable as OIS in Pt, we have considered three more groups, i.e., BIS+BIS, OIS+BIS, and TIS+BIS for a He-pair in Pt.

## 1. A He-pair in *bcc* metals

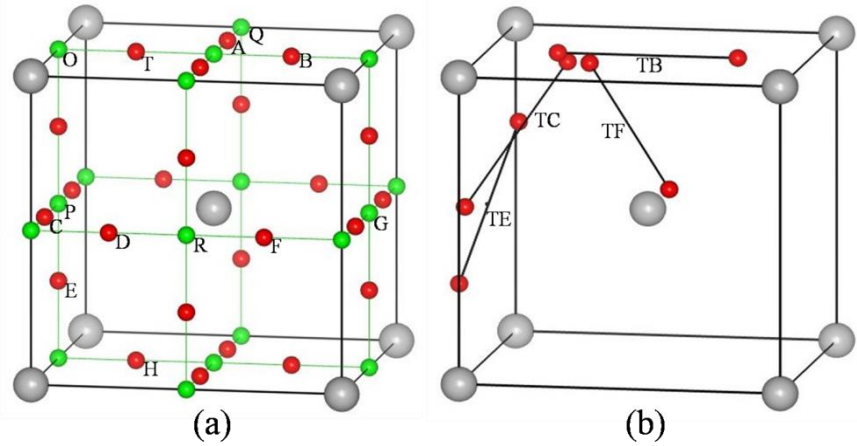

**Fig.3. (a) The initial positions of a He-pair in *bcc* metals, with one He fixed at the position T or O and the other one in a letter-denoted nonequivalent site. Green and red spheres represent OIS and TIS, respectively. (b) The optimized, most stable configurations of a He-pair in Fe. (Those in Mo and W are not shown as they are very similar to the ones in Fe)**

For each *bcc* metal we have calculated the binding energy of a He-pair for 14 configurations, as are illustrated in panel (a) in Fig.3. The four configurations with the strongest binding are depicted in panel (b). We list in Table II the binding energies, initial and optimized inter-atomic distances of a He-pair in 14 configurations. In Fe, Mo and W, configuration TE is always the most stable one, with a He-He binding energy of 0.33, 0.91 and 0.95 eV, respectively. The difference in these binding energies is closely related to the difference in the formation energy of a single interstitial He (c.f. Table I). A larger formation energy of single interstitial He results in stronger attraction of He-pair. Since pairing of two He reduces significantly the area of the He/matrix interface, the reduction in embedding energy of the two He will be more remarkable when each of them has a higher formation energy. In addition to TE, the configuration TC has the same binding energy, 0.33eV, in Fe. Moreover, TB and TF are also only about 0.03 eV less stable than TE. Similarly, TB, TC and TF are also quite stable in Mo and W. Therefore, we have chosen TE as the initial point and combine TB, TC and TF to construct the migration paths when tracing the migration of a He-pair. We want to point out that in our pervious study on He-pair migration in W,[33] we have concluded incorrectly that TF would be optimized into TC. A closer look tells us the optimized configurations of TF and TC are not equivalent.

**Table II. The calculated binding energies, initial and optimized inter-atomic distances of a He-pair in Fe, Mo and W. A He-pair configuration is denoted by two letters denoting the initial positions of the two He, as shown in Fig.3(a), and the initial He-He distance is expressed in terms of lattice constant, *a*.**

| Group | He-pair Configuration | $d_{ini}$ | Fe $d_{opt}$ (Å) | Fe $E_{He2}^b$ (eV) | Mo $d_{opt}$ (Å) | Mo $E_{He2}^b$ (eV) | W $d_{opt}$ (Å) | W $E_{He2}^b$ (eV) |
|---|---|---|---|---|---|---|---|---|
| TIS+TIS | TA | $\sqrt{2}/4a$ | 1.53 | 0.11 | 1.49 | 0.68 | 1.44 | 0.66 |
| | TB | $1/2a$ | 1.59 | 0.31 | 1.54 | 0.88 | 1.50 | 0.89 |
| | TC | $\sqrt{6}/4a$ | 1.64 | 0.33 | 1.58 | 0.88 | 1.53 | 0.92 |
| | TD | $\sqrt{2}/2a$ | 1.67 | 0.26 | 1.70 | 0.40 | 1.65 | 0.35 |
| | TE | $\sqrt{10}/4a$ | 1.60 | 0.33 | 1.55 | 0.91 | 1.51 | 0.95 |
| | TF | $\sqrt{3}/2a$ | 1.63 | 0.29 | 1.56 | 0.90 | 1.52 | 0.93 |
| | TG | $\sqrt{14}/4a$ | TD | TD | TE | TE | TE | TE |
| | TH | $a$ | 2.92 | -0.11 | TB | TB | 3.20 | 0.01 |
| OIS+OIS | OP | $1/2a$ | 1.69 | 0.07 | 1.57 | 0.57 | 1.50 | 0.49 |
| | OQ | $\sqrt{2}/2a$ | 1.53 | 0.11 | 1.49 | 0.68 | 1.44 | 0.66 |
| | OR | $\sqrt{3}/2a$ | 1.61 | 0.18 | 1.65 | 0.39 | 1.56 | 0.34 |
| TIS+OIS | OT | $1/4a$ | TB | TB | TB | TB | TB | TB |
| | OA | $\sqrt{5}/4a$ | TE | TE | TE | TE | TE | TE |
| | OC | $3/4a$ | TB | TB | TB | TB | TB | TB |

## 2. A He-pair in *fcc* metals

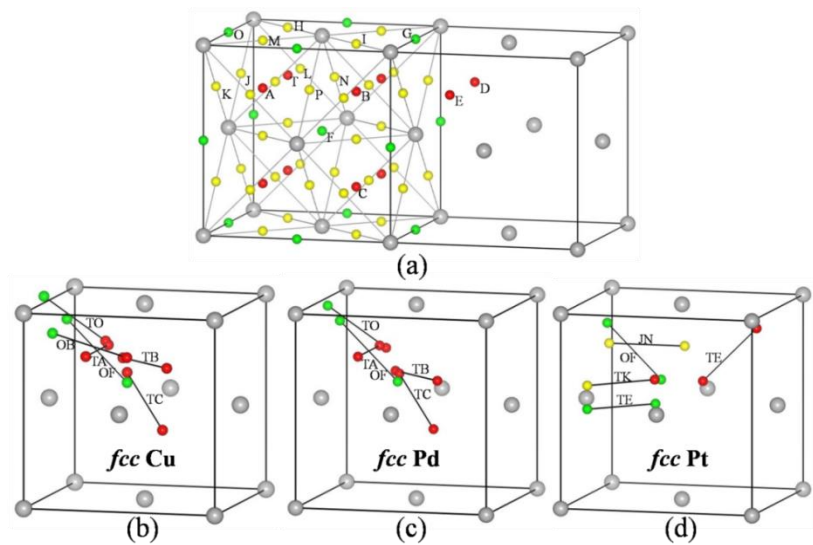

**Fig.4. (a) The initial positions of a He-pair in *fcc* metals, with one He fixed at the position T, O or J, and the other one in a letter-denoted nonequivalent site. The green, red and yellow spheres represented OIS, TIS and BIS, respectively. Panels (b)-(d) display the most stable, optimized configurations in Cu, Pd and Pt.**

In Fig.4(a), we show the initial positions of a He-pair in *fcc* metals, with one He fixed at the position T (a tetrahedral interstitial site), O (an octahedral interstitial site), or J (a bridge interstitial site), and the other one in a letter-denoted nonequivalent site. The most stable, optimized He-pair configurations are displayed in panels (b-d) for Cu, Pd, and Pt, respectively. Apparently, a He-pair shows very similar features in Cu and Pd, but behaves quite differently in Pt. In Table III, we listed the calculated binding energy of 10 He-pair configurations in Cu and Pd. For He-pair in Cu, TO is the most stable configuration with a binging energy of 0.72 eV, consistent with Ref.49. Following TO, there are several other configurations, TA, TB, TC, OF, and OB that are also quite stable. All of these six configurations, with full optimization, are plotted in Fig.4 (b). We have chosen TO as the initial point when tracing the migration of a He-pair.

**Table III. The calculated binding energies, initial and optimized inter-atomic distances of a He-pair in Cu and Pd. A He-pair configuration is denoted by two letters denoting the initial positions of the two He, as shown in Fig.4(a), and the initial He-He distance is expressed in terms of lattice constant, *a*.**

| Group | He-pair Configuration | $d_{ini}$ | Cu $d_{opt}$ (Å) | Cu $E^b_{He2}$ (eV) | Pd $d_{opt}$ (Å) | Pd $E^b_{He2}$ (eV) |
|---|---|---|---|---|---|---|
| TIS+TIS | TA | $1/2a$ | 1.61 | 0.70 | 1.68 | 0.36 |
| | TB | $\sqrt{2}/2a$ | 1.61 | 0.70 | 1.68 | 0.31 |
| | TC | $\sqrt{3}/2a$ | 1.59 | 0.69 | 1.68 | 0.30 |
| | TD | $a$ | 3.80 | 0.08 | 5.93 | -0.24 |
| | TE | $\sqrt{5}/2a$ | 4.00 | 0.10 | 4.37 | 0.06 |
| OIS+OIS | OF | $\sqrt{2}/2a$ | 1.62 | 0.69 | 1.70 | 0.68 |
| | OG | $a$ | 4.40 | 0.02 | 4.55 | -0.06 |
| TIS+OIS | TO | $\sqrt{3}/4a$ | 1.60 | 0.72 | 1.67 | 0.57 |
| | OB | $\sqrt{11}/4a$ | 1.59 | 0.69 | TO | TO |
| | OC | $\sqrt{19}/4a$ | 3.80 | 0.08 | OF | OF |

In Pd, the most stable He-pair configuration is OF, with two He located in a pair of neighboring OIS. The binding energy is 0.68 eV. Configuration TO has a binding energy of 0.57 eV, 0.11 eV lower than OF. We note that configurations TA, TB, and TC are not so close in stability to OF, unlike in the case of Cu. The binding energy of TD is -0.24 eV, an indication of strong repulsion between the two He. This is mainly due to the fact that a single He in OIS is remarkably (0.13 eV) more stable than in TIS. Following the same approach as in the case of Cu, we have chosen OF as the initial point when tracing the migration of a He-pair in Pd.

**Table IV. The calculated binding energy, initial and optimized inter-atomic distances of a He-pair in Pt. For more explanation, see the caption of Table III.**

| Group | Configuration | $d_{\text{ini}}$ (Å) | $d_{\text{opt}}$ (Å) | $E_{\text{He2}}^{\text{b}}$ (eV) |
|---|---|---|---|---|
| TIS+TIS | TA | $1/2a$ =1.99 | 1.57 | 0.51 |
|  | TB | $\sqrt{2}/2a$ =2.81 | 1.57 | 0.59 |
|  | TC | $\sqrt{3}/2a$ =3.45 | 1.52 | 0.49 |
|  | TD | $a$ =3.98 | 5.96 | -0.08 |
|  | TE | $\sqrt{5}/2a$ =4.45 | 1.58 | 1.13 |
| OIS+OIS | OF | $\sqrt{2}/2a$ =2.81 | TE | TE |
|  | OG | $a$=3.98 | 4.62 | 0.01 |
| OIS+TIS | OT | $\sqrt{3}/4a$ =1.72 | 1.56 | 0.98 |
|  | OB | $\sqrt{11}/4a$ =3.30 | 1.57 | 1.11 |
|  | OC | $\sqrt{19}/4a$ =4.34 | 4.09 | 0.40 |
| BIS+BIS | JH | $\sqrt{2}/4a$ =1.41 | TE | TE |
|  | JN | $1/2a$ =1.99 | 1.57 | 1.11 |
|  | JM | $\sqrt{6}/4a$=2.44 | 1.65 | 0.72 |
|  | JP | $\sqrt{2}/2a$ =2.81 | TE | TE |
| OIS+BIS | OJ | $\sqrt{2}/4a$ =1.41 | TE | TE |
|  | OL | $\sqrt{6}/4a$ =2.44 | 1.57 | 1.11 |
| TIS+BIS | TL | $1/4a$ =1.00 | JN | JN |
|  | TK | $\sqrt{5}/4a$ =2.22 | 1.58 | 1.16 |
|  | TI | $3/4a$ =2.99 | OC | OC |

Different from Cu and Pd, Pt has BIS to take into account. We have explored 19 He-pair configurations in Pt. In Table IV, we listed the calculated binding energies, initial and final (optimized) He-He distances of these He-pair configurations. Among them, TK is the most stable one with a binding energy of 1.16 eV. Interestingly, we find that it is spatially very close to configuration TE, but is 0.03 eV more stable than the latter [see Fig.4(d)]. It has to be emphasized that without consideration of BIS, we will fail to discover the TK configuration in Pt. Other He-pair configurations with a binding energy close to TK include OB, JN, and OL, all with a binding energy of 1.11 eV. Obviously they are good candidates to serve as intermediate states on the migration path. Moreover, there are quite a few other configurations that can optimize into either TE or JN. In discussion of He-pair migration below, TE, rather than TK is chosen as the initial point for a migration period, in view of its closeness to the center of high symmetry.

## C. He-pair Migration

### 1. He-pair Migration in *bcc* metals

Thermodynamically, a He-pair has the maximum probability to stay at the most stable configuration. It is therefore advisable to set the most stable configuration as the starting point of a migration step. In Fe, we choose TE as the initial point and combine relatively stable configurations TB, TC, and TF to construct migration paths. First, we have calculated migration barrier from TE to its neighboring TC. This is a combined *rotational-translational* motion. Since it is rather short, in the NEB calculations, we have assumed head-to-head and tail-to-tail correspondence between the initial TE and final configuration TC. The energy barrier for this motion routine is found to be 0.07 eV, as is illustrated in Fig.5(a). We denote this elemental motion as Routine α, and a reversed process from TC to TE is also called Routine α here. The barrier for a He-pair to move from one TC (denoted as TC1) to a neighboring TC (denoted as TC2), is calculated to be 0.04 eV. This motion is nearly *translational*, as is shown below. We name this elemental motion as Routine β, in which TF is the saddle (transition) point, as is demonstrated in Fig.5(b). From TE to its nearest neighbor TB, a He-pair experience a pure rotation. A pure rotation of a He pair in the most stable configuration TE keeps its center of mass in an OIS site. The final state of this rotation is also a TE configuration, which can be viewed as a mirror of the starting geometry. Our first-principles calculation shows the energy barrier for a TE-TB-TE rotation is 0.02 eV. Here we name this *rotational* motion as Routine γ.

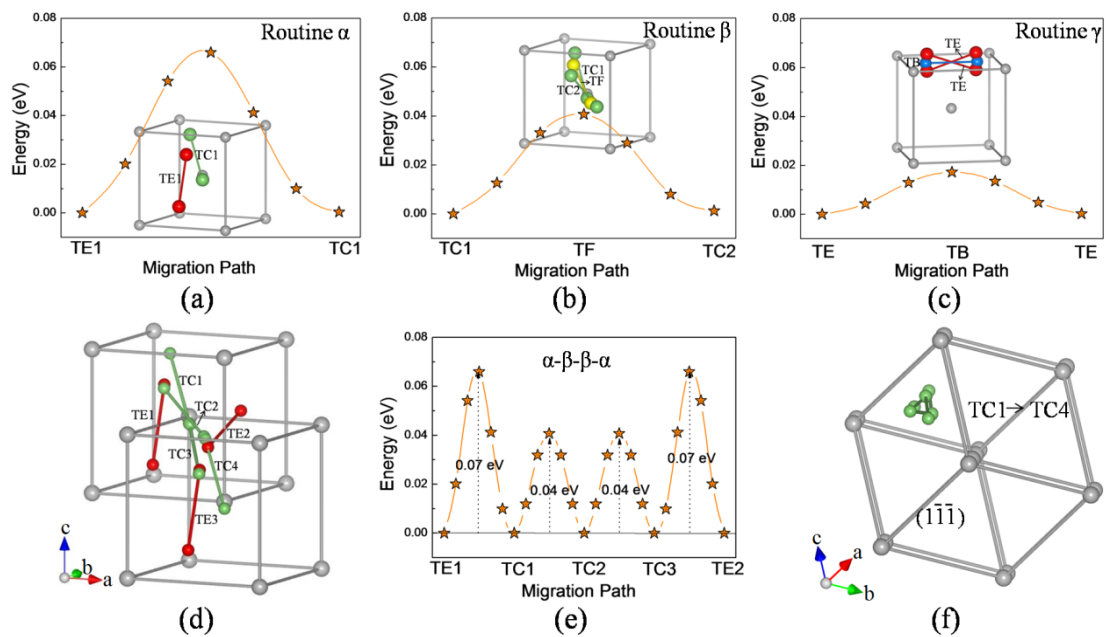

**Fig.5. He-pair migration in *bcc* Fe. (a) A rotational-translational motion from TE1 to its neighboring TC1, Routine α. (b) A translational motion from TC1 to its nearest neighbor TC2, Routine β. (c) A rotational motion from TE to its mirror through TB, Routine γ. (d) TC and TE configurations in the vicinity of TE1. (e) A migration step from TE1 to TE2 constructed by two Routine α and two Routine β. (f) Motions of successive Routine β viewed along [1$\bar{1}\bar{1}$] direction. The red, green, yellow, blue and gray spheres represent configurations TE, TC, TF, TB of a He-pair and Fe atoms.**

The multiple TC and TE configurations located around TE1 are shown in Fig.5 (d). Combining Routines α and β, a He-pair from TE1 not only can move to TE2 through TC1-TC2-TC3 [Fig.5(e)], but also can move to TE3 through successive TC1-TC2-TC3-TC4 [Fig.5(f)]. In the path from TE1 to TE3, the trajectory of successive motions via Routine β is almost a straight line with only slight swaying, which can be seen clearly along the [1$\bar{1}\bar{1}$] direction [Fig.5(f)]. On the other hand, via the same routine, a He-pair from TE2 can move to TE3 through TC3 and TC4 with a barrier of 0.07 eV. Clearly, starting from the most stable configuration TE, a He-pair can migrate to any other TE by at least two Routine α and multiple Routine β, depending on the number of TCs it has to pass by, and the highest migration barrier is 0.07 eV; whereas moving from TC to neighboring TCs, the trajectory is almost a straight line with a barrier of 0.04 eV.

The migration of He-pair also starts from the most stable configuration TE in Mo. Different from that in Fe, the second stable configuration in Mo is TF, not TC, therefore, Routine β will start from configuration TF, using TC as the transition point. Similar to the case of Fe, the migration barriers and motion Routines α, β and γ of a He-pair are shown in Fig.6(a), (b) and (c), respectively. From TE1 to its neighboring TF1, the energy barrier is found to be about 0.07 eV, the same as in Fe. And from TF1 to TF2, it is as small as 0.02 eV. The energy barrier for Routine γ, a TE-TB-TE rotation is 0.03 eV. From TE1 to TE2, a He-pair will pass through TF1 and TF2 [Fig.6(e)], and from TE1 to TE3 it will pass through successively configurations TF1-TF2-TF3-TF4 [Fig.6(f)]. Successive Routine β motion, viewed along the [1$\bar{1}\bar{1}$] direction is depicted in Fig.6 (f), a path even more close to straight line than that in Fe. The reason for a straighter trajectory is that the He-He distance in Mo is shorter than in Fe. Similarly, starting from the most stable configuration TE, a He-pair can migrate to any other TE by at least two Routine α and multiple Routine β, depending on the number of TCs it has to pass by, and the highest migration barrier is 0.07 eV. The energy barrier for Routine γ, a TE-TB-TE rotation is 0.03 eV.

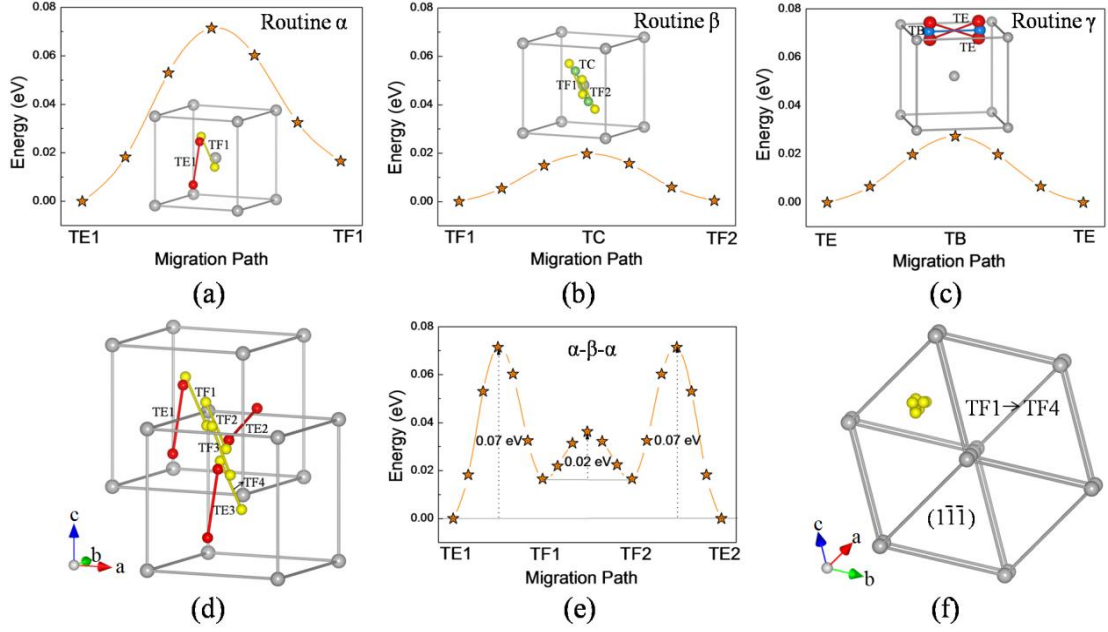

**Fig.6. He-pair migration in *bcc* Mo. (a) A rotational-translational motion from TE1 to its neighboring TF1, Routine α. (b) A translational motion from TF1 to its nearest neighbor TF2, Routine β. (c) A rotational motion from TE to its mirror through TB, Routine γ. (d) TF and TE configurations in the vicinity of TE1. (e) A migration step from TE1 to TE2 constructed by two Routine α and one Routine β. (f) Motions of successive Routine β viewed along [$\bar{1}$11] direction. The red, yellow, green, blue and gray spheres represented configurations TE, TF, TC, TB of a He-pair and Mo atoms.**

An interstitial He-pair in W behaves similarly to the one in Mo. We plot three kinds of migrations: Routine α in Fig.7(a), Routine β in Fig.7(b) and Routine γ in Fig.7(c). The calculated migration barriers are 0.08 eV, 0.01 eV and 0.06 eV, respectively. Therefore the migration barrier is 0.08 eV, significantly lower than the value we obtained in a previous work, 0.30 eV.[33] One of the reasons we obtained a much higher migration barrier in the previous work, as mentioned above, is that we have by mistake thought TF would be optimized into TC, therefore, we did not take TF into account when constructing the migration routines. The other reason is that we have been misled by the closeness of TE and TC configurations, and assumed by mistake that TE to TC is rather a transformation than a real motion, a piece of migration path.

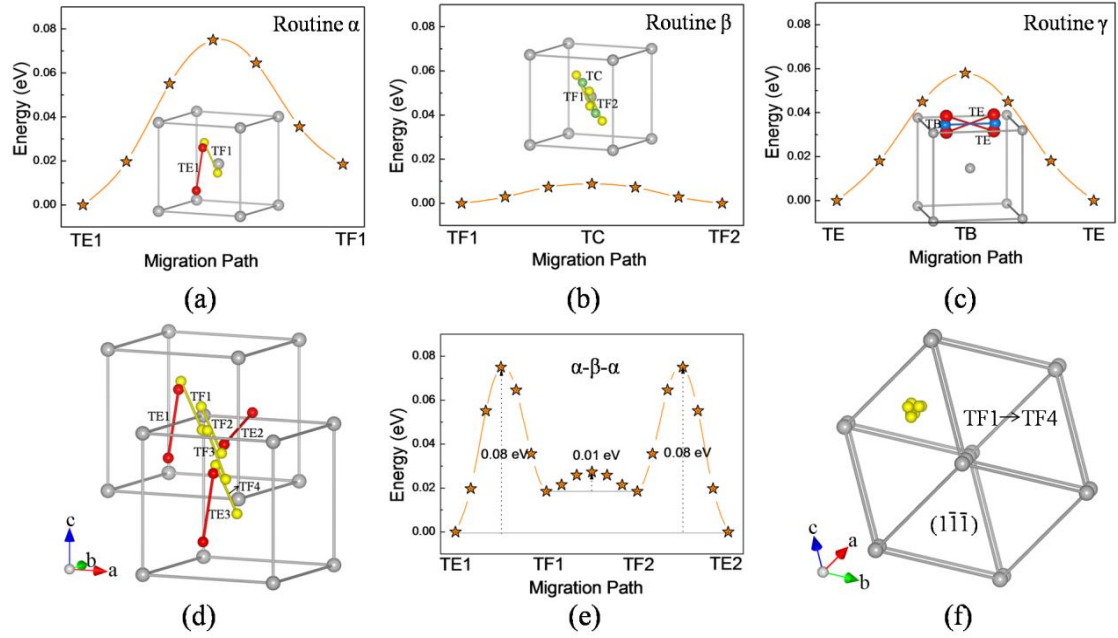

**Fig.7. He-pair migration in *bcc* W. For explanation, see the caption of Fig.6.**

With the elemental motion routines α, β, and γ available, a He-pair can migrate at will from one TE, its most stable configuration, to another. The migration path is certainly irregular in long-range. However, within the scale of one or a couple of lattice constants, the trajectory of a dancing He-pair can be possibly regular. We now construct *full* migration steps with combinations of Routines α, β and γ. By *full* we mean the center of mass of a He pair move from position $\vec{P}$ (an OIS, for instance) to $\vec{P} + l\vec{a} + m\vec{b} + n\vec{c}$, where $\vec{a}, \vec{b}$, and $\vec{c}$ are unit vectors with a length of lattice constant. In Fig.8(a), we can employ migration process from TE1 to the nearest neighboring TE2 [α-β-α, panel (e) in Fig.5-7.] to realize a full migration step along [100] direction. The path is zigzag in three-dimension. Notice that to guide the eye, we make the head He red and the tail He green, and number sequentially the pair in different positions. Also, only some intermediate pairs are numbered, bonded, and have the normal size, other points are displayed using small circles. The same method is also used in following other kind of motions.

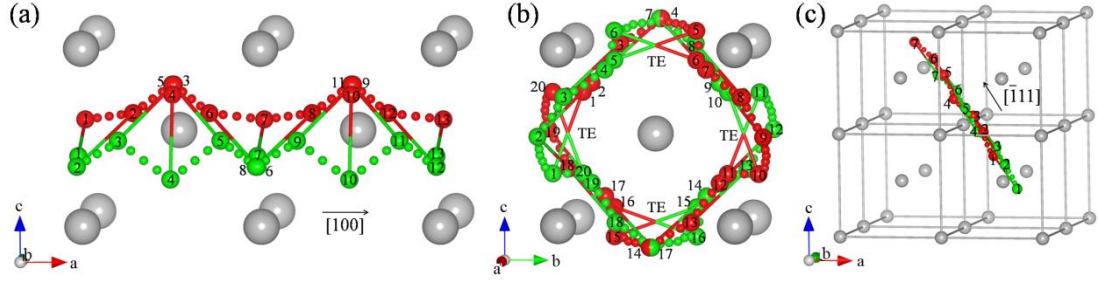

**Fig.8. Migration of a He-pair in *bcc* metals Fe, Mo, and W. (a) Zigzag three-dimensional motion along [100]; (b) Wavy circular motion in (100); (c) Swaying one-dimensional motion along [$\bar{1}$11]. To guide the eye, we make the head He red and the tail He green, and number sequentially He-pair in different positions. In addition, only some intermediate pairs are numbered, bonded, and have the normal size, other points are displayed using small circles.**

On the other hand, we combine migration process TE2→TE3 (the second nearest neighboring TE configuration) and pure rotation TE→TB→TE (Routine γ) to make a half circle through α, β and γ. We note that this displacement can be done equally if the pair chooses the other half of this circle. We call this a wavy circular motion in (100). And likewise, if we connect migration process from TF to neighboring TF (for Mo and W) or TC to neighboring TC (for Fe), i.e., β-β-…-β, we will obtain a swaying one-dimensional motion along [$\bar{1}$11] direction. Care should be taken when we view this translational motion. Although configuration TC (in Fe) or TF (in Mo and W) can move forward along [$\bar{1}$11] direction with lower barrier, an additional barrier is needed to promote a He-pair from the most stable configuration TE.

In addition, we have also constructed other migration paths making use of configuration TA, but they all have much higher energy barriers than what we obtain here using α, β, and γ combinations. Thus we are now confident that the energy barrier of 0.07, 0.07, and 0.08 eV in Fe, Mo and W, respectively.

## 3. He-pair migration in *fcc* metals

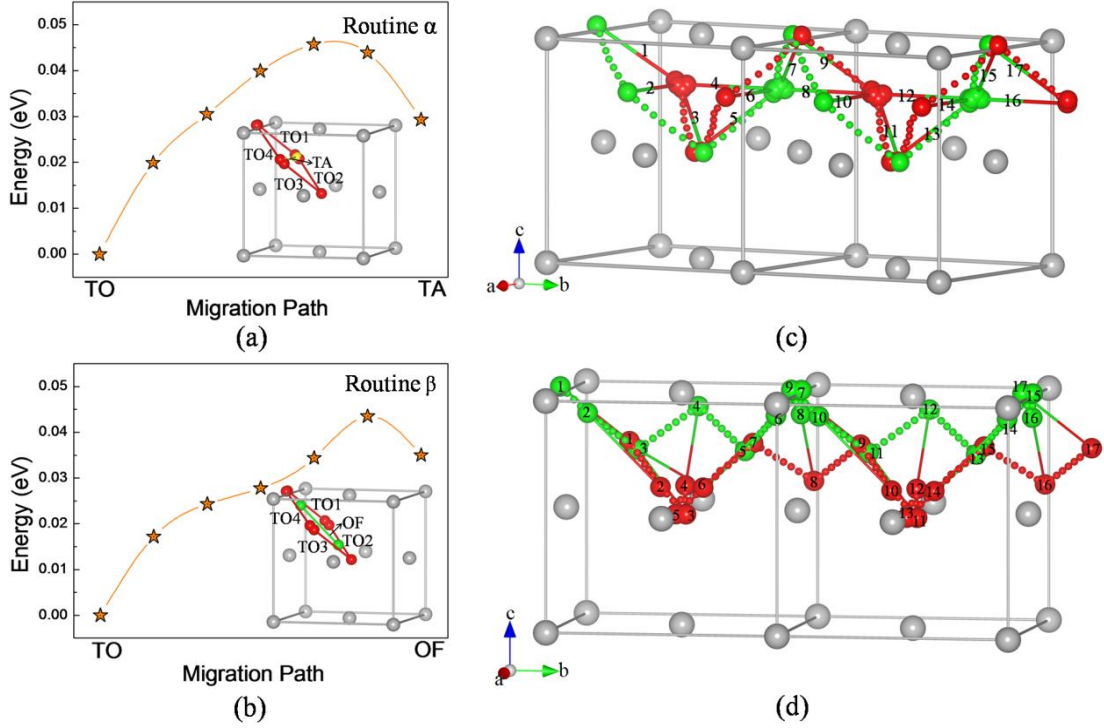

**Fig.9. He-pair migration in *fcc* Cu. (a) TO→TA, Routine α. (b) TO→OF, Routine β. Inset: the red, yellow, green and gray spheres represent configurations TO, TA, OF and Cu atoms. (c) Migration trajectories composed of Routine α; (d) migration trajectories composed of Routine β. We make the head He red and the tail He green, and number sequentially He-pair in different positions. In addition, only some intermediate pairs are numbered, bonded, and have the normal size, other points are displayed using small circles.**

We can learn from Table III that the most stable configuration of a He-pair is TO, optimized from a pair with one in OIS and the other in its nearest neighbor TIS. Several other configurations such TA, TB, and TC (TIS+TIS combination) and OF (OIS+OIS combination) have a binding energy just slightly lower than TO. Thus, we choose TO as the initial point and consider relatively stable configurations TA, TB, TC, and OF to construct migration paths. Since the migration path for a single He is TIS-OIS-TIS [Fig.2(a)], we are curious about the motion of a He-pair from TO to TA [see Fig.4(a)], during which only one He moves. Our calculation shows that the energy barrier for this elemental motion is 0.046 eV. We name this motion as Routine α and display it in Fig.9(a). With one He nearly standing still and the other rotating [TO1-TA], this motion combines rotation and translation. In the inset of Fig.9(a), we can also see that the configuration TO2, TO3 and TO4are equivalent to TO1. It follows immediately that the motion from TA to other TOs should also be 0.05 eV, thus it is named Routine α. This means that through two Routine α, a He-pair TO can move to its neighboring equivalent site. On the other hand, in Fig.4(a), we have noticed another stable configuration, OF, similar as TA relative to four TO configurations. This is a strong hint that the motion from one TO to its neighboring TO probably has a low barrier if the low energy OF serves as the transition state. Expectedly, our calculations yield an energy barrier of 0.05 eV, nearly the same as Routine α. We name this motion as Routine β and display it in Fig.9(b).

In panels (c) and (d) of Fig.9, we construct two *full* migration steps using Routine α and Routine β respectively. Since the energy barriers for Routines α and β, both are 0.05 eV, are only slightly higher than the energy increase from TO to TA (0.02 eV) and that from TO to OF (0.03 eV), we believe that the migration paths we constructed using Routines α and β are presumably the lowest, i.e., the true paths for a He-pair in Cu. It is worth noting that the migration barrier for a He-pair is lower than that for a single He. We will discuss this intriguing phenomenon in the following section.

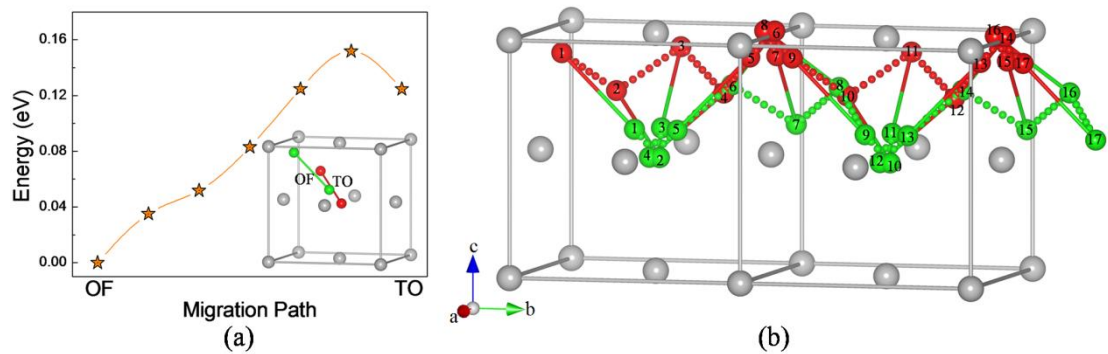

**Fig.10.He-pair migration in *fcc* Pd. (a) The calculated energy barriers of OF→TO. (b) Trajectories on the basis of OF→TO routine for He-pair in Pd. We make the head He red and the tail He green, and number sequentially He-pair in different positions. In addition, only some intermediate pairs are numbered, bonded, and have the normal size, other points are displayed using small circles.**

Unlike in Cu, the most stable configuration of a He-pair in Pd is OF, optimized from a pair with both He in OIS. However, we have to point out that after optimization, none of the two He is in OIS, as is clearly seen in Fig.4(c). And more importantly, we see from Table III that there is only one other configuration, TO, has a binding energy (0.57 eV) close to OF (0.68 eV). Based on the knowledge we have just obtained for the Cu system, we calculated only the migration from configuration OF to TO (Routine β in Cu). In Fig.10(a), we plot the migration path, which has a barrier energy of 0.15 eV. In panel (b) we construct a *full* migration step using Routine β, which has a very similar appearance as the one shown in Fig.9(d). Different from the Cu case, the migration barrier for a He-pair in Pd is equal to that for a single He.

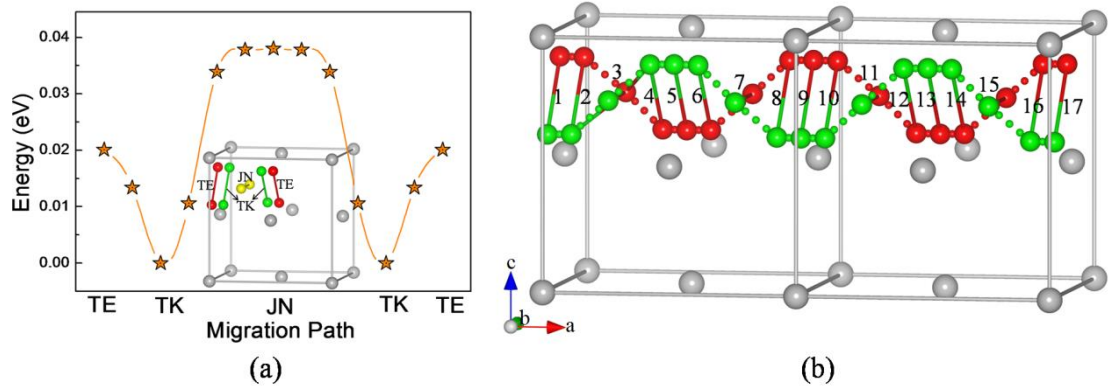

**Fig.11. He-pair migration in *fcc* Pt. (a) The energy barriers of TE→JN→TE. (b) Full migration trajectories along $\vec{a}$ on the basis of TE→JN→TE routine. The metastable intermediate configurations are marked, bonded and have the normal size. Other points on trajectory are plotted by using smallest spheres.**

The fact that a BIS is almost as stable as an OIS makes Pt much different from Cu and Pd in the sense of He migration. Similar to the single He case, a He-pair could also squeeze through BIS on their way from an OIS to a neighboring OIS. Since TE is closer to high symmetry point, we prefer to use TE, rather than TK, as the initial point for a migration step. As displayed in Fig.10(a), we have calculated the migration path TE→JN→TE along <001>, and the calculated energy barrier is 0.04eV. In panel (b) we construct a *full* migration step, whose trajectory looks like the double helix of DNA.

## IV. DISCUSSION

Having described the calculated results of the migration barriers and paths for both single He and He-pair in each metal, we now recap the main numerical results in Table V. There are several observations deserving further discussion.

**Table V. The formation energy of an interstitial He atom ($E_{\text{He1}}^{\text{f}}$), the energy difference between the most and second stable interstitial sites among OIS, TIS, and BIS ($\Delta E_{\text{He1}}^{\text{f}}$), the migration barrier for a single He ($E_{\text{He1}}^{\text{m}}$); the distance ($d_{\text{He-He}}$), binding energy ($E_{\text{He2}}^{\text{b}}$) between the two He in a pair, the energy difference between the most and second stable He-pair configurations ($\Delta E_{\text{He2}}^{\text{f}}$), the migration barrier ($E_{\text{He2}}^{\text{m}}$) for He-pair in *bcc* Fe, Mo, and W, and *fcc* Cu, Pd, and Pt.**

| Matrix | Fe | Mo | W | Cu | Pd | Pt |
|---|---|---|---|---|---|---|
| $E_{\text{He1}}^{\text{f}}$ (eV) | 4.61 | 5.36 | 6.14 | 4.00 | 3.60 | 4.83 |
| $\Delta E_{\text{He1}}^{\text{f}}$ (eV) | 0.20 | 0.17 | 0.20 | 0.02 | 0.13 | 0.03 |
| $E_{\text{He1}}^{\text{m}}$ (eV) | 0.06 | 0.06 | 0.06 | 0.08 | 0.15 | 0.03 |
| $d_{\text{He-He}}$ (Å) | 1.60 | 1.55 | 1.51 | 1.60 | 1.70 | 1.58 |
| $E_{\text{He2}}^{\text{b}}$ (eV) | 0.33 | 0.91 | 0.95 | 0.72 | 0.68 | 1.16 |
| $\Delta E_{\text{He2}}^{\text{f}}$ (eV) | 0.00 | 0.01 | 0.02 | 0.02 | 0.11 | 0.03 |
| $E_{\text{He2}}^{\text{m}}$ (eV) | 0.07 | 0.07 | 0.08 | 0.05 | 0.15 | 0.04 |

### A. The site-preference of He in Cu is different from in Pd and Pt

A single interstitial He prefers TIS in all three *bcc* metals and OIS in Pd and Pt, however, the most stable interstice in Cu is TIS. Why is the site-preference in Cu different from in Pd and Pt? Here, we introduce a new concept, *point elastic constant*, which is defined as the tensile constant of a regular polyhedron in a crystal under local radial stress originated from the center of this polyhedron. Since OIS and TIS in *fcc* metals exhibit high symmetry under tension and compression, it is reasonable to assume that the local polyhedron is isotropic. The embedding energy of a He atom in an interstitial site can then be decomposed phenomenologically into two contributions: one is the compression from a homogeneous electron gas, and the other from the compression imposed by surrounding matrix atoms. The energy cost to embed a He into an electron gas is in proportion to the electron density, $\rho_0$;[50] and the energy cost to push the surrounding atoms outward is approximated by an elastic energy which can be evaluated using the *point elastic constant* for a particular interstitial site. Thus, we can estimate the embedding energy of He into different interstices with

$$\Delta E \cong \alpha \rho_0 + \frac{1}{2} k \Delta d^2 \quad (3)$$

Here, $\alpha$, $\rho_0$, $k$, $\Delta d$ are respectively the scaling factor of electron density, electron density of the interstice, *point elastic constant* of interstice, and distance change between interstice and the nearest neighboring matrix atom. The *point elastic constant*, $k$, can be obtained by the second derivative that gives the changes in energy with respect to the variation of distance ($\pm 0.69\%$ and $\pm 1.38\%$ of the lattice constant) between interstices and the nearest neighboring matrix atoms in the absence of embedded He. In the calculations, the nearest neighboring matrix atoms of an interstice were fixed, while all other metal atoms were relaxed. The *point elastic constant* of OIS and TIS in Cu, Pd and Pt are listed in Table VI.

**Table VI. Electron density $\rho_0$ ($e$/a.u.$^3$) and point elastic constant $k$ (eV/Å$^2$) of TIS and OIS interstices in Cu, Pd and Pt.**

| Matrix | OIS | | | TIS | | |
|---|---|---|---|---|---|---|
| | $\rho_0$ | $k$ | $\rho_0 \times k$ | $\rho_0$ | $k$ | $\rho_0 \times k$ |
| Cu | 0.021 | 58.881 | 1.237 | 0.031 | 37.374 | 1.159 |
| Pd | 0.017 | 70.889 | 1.205 | 0.028 | 46.245 | 1.295 |
| Pt | 0.019 | 91.383 | 1.736 | 0.035 | 58.683 | 2.054 |

It is seen that the electron density at OIS is always lower than at TIS in Cu, Pd, and Pt, while the *point elastic constant* is just on the opposite. Here, we tentatively employ a quantity, $\rho_0 \times k$, to represent the competition of the two contributions to the embedding energy for He in OIS and TIS. Although the electron density at TIS is higher than that at OIS, its effect can be over compensated by the lower *point elastic constant* at TIS than at OIS, in the case of Cu. This is not found, however, in Pd and Pt. As a consequence, a He atom has a smaller embedding energy in TIS than in OIS in Cu; whereas in Pd and Pt the OIS is more stable than TIS for He.

It is worth noting that not only in Cu, the TIS was also found to be more stable than OIS in Ni.[25] We have made additional calculations for positioning of He in *fcc* lattice of Co, Mn, and Be, and find that TIS gets more and more stable with decreasing lattice constant. In a small lattice, the gradient of electron charge density at the interstices will also be reduced. Therefore, interstitial point elastic character will play a more significant role in site preference of the embedded He.

**B. Helium migration is quite similar in different *bcc* metals**

Interestingly, the migration barrier for both single He and He-pair is quite similar in Fe, Mo and W. Firstly, within the precision of 0.01 eV, the migration barrier of a single interstitial He is 0.06 eV, the same in all three metals, and the migration path is identical as well [c.f. Fig.2(a)]. This is primarily due to the same site-preference order and the closeness of formation energy difference between TIS and OIS (0.17 eV for Mo and 0.20 eV for Fe and W). Moreover, the migration barrier for a He-pair is also very similar, 0.07 eV in Fe and Mo and 0.08 eV in W. The migration of a He-pair is dominated by energy differences between the most stable He-pair configurations, which are very small in Fe, Mo, and W. The most stable He-pair configurations, TB, TC, TE and TF, are all TIS-TIS type, and determine the migration paths. Since He-pair configurations are associated with the stability of single He in different interstices, the similarity in their migration barrier in different metals mainly stems from the same site-preference for single He. We stress that both the energy barrier and the migration path for a single interstitial He are rather independent of the solution energy of He, and those for a He pair are rather independent of the paring energy of He. Therefore, we can anticipate that the energy barrier for larger He clusters might also be similar in different *bcc* metals. Such a prediction, of course, calls for further *ab initio* investigations.

## C. Helium migrates dissimilarly in different *fcc* metals

Unlike in *bcc* metals, the migration of both singe He and He-pair in *fcc* metals is quite material-dependent. The migration barrier for a single He is 0.08, 0.15, 0.03 eV in Cu, Pd and Pt, and the motions are via TIS→OIS→TIS, OIS→OIS, and OIS→OIS respectively. Such significant differences are essentially due to dissimilarity in the positioning of interstitial He, which is TIS in Cu and OIS in Pd and Pt. There is a strong correlation between the migration barrier and the difference in formation energy between the most and second most stable interstice, $\Delta E_{\text{He1}}^{\text{f}}$, as shown in Table V. The small difference in stability between TIS and OIS in Cu, 0.02 eV, leads to a low migration barrier in Cu. On the other hand, the low migration barrier (0.03 eV) in Pt comes with the existence of a relatively stable interstice, BIS. As for He-pair migration, our exhaustive search for stable He-pair configurations enables us to construct reliable migration paths. Owning to metastable BIS, the stable configurations and stability-order of a He-pair in Pt are different from in Cu and Pd. The most probable migration path in Pt is somewhat like a double helix of DNA, distinguishing from that in Cu and Pd. One of two most probable migration paths in Cu is similar to the most probable path in Pd, but the migration barrier (0.05 eV) is much smaller than the latter (0.15 eV), a result of a small difference in the formation energy of heat TIS and OIS. Similarly, due to the closeness of formation energy difference between OIS and BIS, the energy barrier for He-pair migration is very small, only 0.04 eV in Pt.

## D. Migration energy of a He-pair is close to a single He

Our first-principles calculations demonstrate unambiguously that the migration barrier for an interstitial He-pair is about equal to that for a single interstitial He in all metals except Cu, in which a He-pair can even move faster than single He. This is rather counterintuitive. Intuitively, it is expected that a larger strain field leads to a higher diffusion energy barrier. However, the energy barrier is determined by the stability difference between He cluster at the minimum point and the saddle point, not by the absolute formation energy of impurities. Similar observation was reported in reference [51] that larger solute atoms can move faster than smaller ones.

To elucidate the underlying force responsible for associated He-pair migration, we have investigated the interaction between two He atoms against their inter-atomic distance. Here, we have studied the most stable He-pair configuration in each metal. Base on the knowledge we obtained from our recent work on W,[33] we know that He-He binding in other low-energy configurations is very similar to this one. One He-pair defines a straight line and we study an ideal motion of the two He atoms on the line. We plot in Fig.12 the calculated binding energy of He-pair in *bcc* and *fcc* metals as a function of varying inter-atomic distance, in reference to the most stable single interstitial He. Negative values mean He-He attraction. As was in done reference [33], we have fixed the positions of two He and allowed all the other metal atoms to relax upon energy minimization.

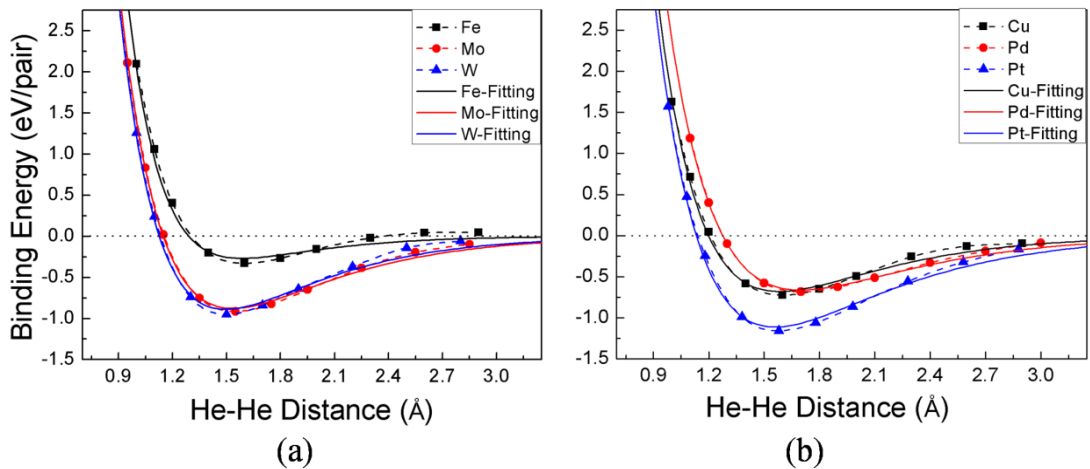

**Fig.12. Binding energy of an interstitial He-pair with a varying inter-atomic distance in *bcc* Fe, Mo and W (a) and *fcc* Cu, Pd and Pt(b), in reference to the most stable single interstitial He. Solid curves are fitted Morse Potential and the dotted lines connecting the calculated values are used to guide the eye.**

Obviously, we can find that the interaction of two interstitial He atoms in all metals resembles very much to typical chemical bonds. Starting from far apart, the two He atoms attract each other until at about 1.5Å, they show strongest binding. When the distance decreases down below 1.20 Å, they start to exhibit repulsion. We make an attempt to describe the chemical bonding-like interaction between He-pair in metals using Morse potential. Morse potential can give a reasonable short-range repulsion and long-range attraction and have been used as to calculate the energy of vaporization, the lattice constant, and the compressibility in cubic metals.[52] The Morse potential takes the form as

$$V(r) = D_e\left(e^{-2a(r-r_e)} - 2e^{-a(r-r_e)}\right) \qquad (4)$$

Where $r$ is the distance between the interacting atoms, $r_e$ is the equilibrium bond distance, $D_e$ is the well depth, and $a$ controls the width of the potential (the smaller the $a$ is, the larger is the well). This form approaches zero at infinite $r$ and equals $-D_e$ at its minimum, i.e. attraction of $D_e$ at $r=r_e$. We only consider the most stable He-pair configuration, whose probability is the highest. The fitting Morse potentials are plotted as solid curves in Fig.12. The fitting empirical parameters $D_e$, $a$, and $r_e$ for each metallic system are shown in Table VII.

**Table VII. Fitting empirical parameters $D_e$, $a$, and $r_e$ in Morse potential function in various metals. The binding energy of He-pair given by DFT is listed as well.**

| Matrix | $D_e$ (eV) | $a$ (Å$^{-1}$) | $r_e$ (Å) | $E_b$ (eV) |
|:---:|:---:|:---:|:---:|:---:|
| Fe | 0.28 | 2.33 | 1.58 | 0.33 |
| Mo | 0.88 | 1.80 | 1.53 | 0.91 |
| W | 0.89 | 1.89 | 1.49 | 0.95 |
| Cu | 0.68 | 1.81 | 1.58 | 0.72 |
| Pd | 0.66 | 1.64 | 1.69 | 0.68 |
| Pt | 1.11 | 1.62 | 1.56 | 1.16 |

The consistency between DFT results and the fitting Morse potential is quite impressive, strongly indicating that the interaction between He-pair in metals could be well described by chemical bonding. The fitting parameters could be used in future simulations on larger He clusters in metals. In all metals, the He-He attraction is strong enough to hold the pair upon passing migration barriers to accomplish 3D rotational-translational motions. The strong attraction suggests another He embedding can ease the strain field and induced lattice distortion compared with simple sum of two He atoms. More important is that the strong attraction is not very short-ranged, and this is very crucial in holding the pair against thermal disturbance.

## V. CONCLUSIONS

In summary, we have performed density functional theory calculations to study the self-trapping and migration of an interstitial He-pair in *bcc* (Fe, Mo and W) and *fcc* (Cu, Pd and Pt) metals. By exhaustive search we have determined the most stable configurations of an interstitial He-pair in each metal. To reveal its migration paths and energy barriers, we decompose its motion into rotational, translational, and rotational-translational routines, which are then used to construct the full migration steps. Our first-principles calculations demonstrate that the migration trajectories and barriers are determined predominantly by the relatively stable He-pair configurations, which, in turn, are strongly dependent on the formation energy of a single He in different interstices. The energy barriers for He-pair migration are only marginally higher than those for a single interstitial He in all studied metals except for Cu, in which a He-pair can even move faster than a single He. The associative motions of a He-pair are guaranteed by the *chemical bonding*-like strong He-He interaction in metals, which can be well modeled by the Morse potential.

To understand why He prefers TIS to OIS in Cu, we have introduced a concept of *point elastic constant* to describe the degree of difficulty for an interstitial atom to push outward the surrounding matrix atoms. Although the electron density at TIS is higher than that at OIS, its effect can be over compensated by the lower *point elastic constant*, unlike in Pd and Pt. Another interesting discovery is that the migration barrier for both single He and He-pair is quite similar in *bcc* metals. For single He, the similarity is primarily due to the same site-preference order and the closeness of formation energy difference between TIS and OIS. The migration of a He-pair is dominated by energy differences between the most stable He-pair configurations, which are determined by the stability of single He in different interstices.

Unlike in *bcc* metals, the migration of both singe He and He-pair in *fcc* metals is quite material-dependent. Such significant differences are essentially due to dissimilarity in the positioning of interstitial He, which is TIS in Cu and OIS in Pd and Pt. The small difference in stability between TIS and OIS in Cu, 0.02 eV, leads to a low migration barrier in Cu. On the other hand, the low migration barrier (0.03 eV) in Pt comes from the existence of a relatively stable interstice, BIS.

We believe the construction of efficient inter-atomic potentials should involve the migration barrier not only for a single He atom, but also that for a He pair or even a He-trimmer which can now only be determined by first-principles DFT, but not experimental techniques at present. Therefore, our density functional theory results on the migration of single He and He pair in both *bcc* and *fcc* metals not only provide a starting point for future first-principles explorations on the growth and migration of larger He clusters in metals, but also serve as touchstones to test the rationality of pair-potentials or many-body-potentials developed for the He-in-metal systems, without which the detailed knowledge of He bubble growth would be out of reach.